\documentclass[twocolumn,times,tighten]{aastex63}
\usepackage{natbib}

\accepted{by ApJ, 2021 Nov 10}
\shorttitle{Locations of Sgr B2 and Star $\iota$}

   \shortauthors{Oka \& Geballe}
\begin{document}

\title{The Central 300\,pc of the Galaxy Probed by Infrared Spectra
  of H$_3^+$ and CO: \\ III.  Locations of Sgr B2 and Star $\iota$}

\author{Takeshi Oka}
\affiliation{Department of Astronomy and Astrophysics and
  Department of Chemistry, the Enrico Fermi Institute,
  University of Chicago, Chicago, IL 60637 USA;
  t-oka@uchicago.edu}
\author{T. R. Geballe} \affiliation{Gemini Observatory/NSF's NOIRLab, Hilo, HI 96720 USA}
\%================================================================
\vspace{4mm}
\begin{abstract}
 Using a simple relation between the radial expansion velocity of diffuse gas in the Central Molecular Zone (CMZ) of the Galaxy and its distance from Sgr~A$^\ast$ we estimate the physical depths within the CMZ of Star $\iota$ (2MASS $J$17470898-2829561) and two Sgr B2 far-infrared continuum sources with respect to the location of Sgr A$^\ast$. To do this we use velocity profiles of  mid-infrared absorption lines of H$_3^+$ and of far-infrared absorption lines of H$_2$O$^+$, OH$^+$, and $^{13}$CH$^+$. The distances to Star $\iota$ and to the Sgr B2 sources are found to be $\sim$90 pc greater than the distance to Sgr~A$^\ast$. Our conclusion that Sgr B2 lies toward the rear of  the CMZ is contrary to many previous models in which it has been placed shallower than Sgr~A$^\ast$.
\end{abstract}
\section{INTRODUCTION}

The Galactic center's (GC's) giant molecular cloud complex Sgr B2 has played a major role in the discoveries of interstellar molecules. Beginning with the discovery of H$_2$O \citep{che69} about 70 molecules have been first observed in Sgr B2 \citep{woo21} in emission in spite of its large distance and consequent high dilution factor relative to nearer molecular clouds. 

There seems to be no consensus on the physical depth of Sgr B2 within the Central Molecular Zone (CMZ) of the Galaxy \citep{mor96}, a roughly disk-like region of radius  $\sim$150 pc and thickness many tens of pc. It is not even known with certainty whether Sgr B2, which is approximately a distance of 100 pc on the plane of the sky from the supermassive black hole Sgr~A$^\ast$, generally considered to lie at the very center of the CMZ (herein we use a GC distance of 8 kpc), is to the front or the rear of Sgr~A$^\ast$ as viewed from the Sun. Initially \citet[][see their Figure 10]{roh82}  placed Sgr B2 far behind Sgr~A$^\ast$ in their Galactic center model inferred from observations of HI emission. Later \cite[][Figure 10]{sof95} positioned Sgr B2 as a part of his Arm I about 90 pc in front of Sgr~A$^\ast$ from extensive analysis of the CO emission in the GC surveyed by \citet{bal87,bal88}. \citet[][Figure 11]{saw04} placed Sgr B2 slightly to the rear of Sgr~A$^\ast$ from analyses of CO and OH spectra. \citet{ryu09} concluded Sgr B2 is on the near side from their X-ray study. On the other hand, both \citet[][Figue 21]{bal10} and \citet[][Figure 5]{mol11} clearly placed Sgr B2 to the rear of Sgr~A$^\ast$ from their mappings of the far-infrared dust continuum. A measurement of the relative proper motions of Sgr B2 and Sgr~A$^\ast$, reported in a paper on the trigonometric parallax of Sgr B2 by Reid et al. (2009), led those authors to estimate that Sgr B2 is 130 $\pm$ 60 pc in front of Sgr~A$^\ast$, which might put it outside of the CMZ. \citet[][Figure 6]{kru15} placed Sgr B2 38 pc in front of Sgr~A$^\ast$ on their proposed noncrossing orbit. \citet[][Figure 9]{rid17} also have Sgr B2 in front of Sgr~A$^\ast$, but by about 80 pc, which is near the front of the CMZ. Most recently \citet{arm20} concluded that Sgr B2 is on the far side of the CMZ based on the placement of Sgr B2 along Arm I by \citet{sof95,sof17}, but adopting the alternative interpretation of Ridley et al. that Arm I is the far spiral arm and Arm II is the near spiral arm.

Among the sightlines toward 18 stars which were used as background sources for obtaining the infrared spectra of H$_3^+$ presented in \citet{oka19}, hereafter Paper I, and in \citet{oka20}, hereafter Paper II, the sightline toward the star that these authors labeled Star $\iota$ (2MASS $J$17470898-2829561), located 84.5 pc east of Sgr~A$^\ast$,  stands out as having unique velocity profiles of lines of H$_3^+$, extending to much larger positive velocities than the profiles toward any of the other stars. Star $\iota$, selected in a search for bright stars with smooth continua \citep{geb19}, at Galactic longitude 0\fdg5477, is located between the nominal centers of two giant molecular clouds, Sgr B2 at longitude 0\fdg6667 and Sgr B1 at longitude 0\fdg5059 and is about 17 pc west of the center of Sgr B2. The unprecedented strengths of its  H$_3^+$ and low $J$ overtone CO lines strongly suggest that it is physically associated with Sgr B \citep{geb10}. Star $\iota$, two stars, $\lambda-$ and $\lambda-+$ to the east of it, together with Star $\alpha$ and its nearby neighbors $\alpha+$ and $\beta$, located near the western edge of the CMZ (see Table 2 of Paper I) have been the most important stars for elucidating the gas dynamics derived in Paper II (see its Figure 11 for their longitudinal locations in the CMZ). \citet{geb21} provide details of spectra toward and of Star $\iota$. Since this star is close to Sgr B2, the sightline toward it may pass through some of the same clouds as sightlines toward the bright far-infrared sources near the center of Sgr B2.

In this paper we introduce a new method of determining the locations within the CMZ of Star $\iota$ and Sgr B2 in the CMZ using velocity profiles of the infrared vibration-rotation lines of H$_3^+$ and far-infrared rotational lines of H$_2$O$^+$, OH$^+$, and CH$^+$.

\section{Radially Expanding Diffuse Gas Observed in the Infrared Spectrum of H$_3^+$}

In Paper I we demonstrated that the most of the volume of the CMZ contains warm ($T$$\sim$200 K) and diffuse ($n$$\sim$50 cm$^{-3}$) gas, where $n$ is the particle number density. In Paper II we concluded that this gas is moving radially outward from an origin near Sgr~A$^\ast$  at speeds of up to $\sim$150 km s$^{-1}$ with the outer circumferential surface of the expansion, located $\sim$150 pc from the center of the CMZ, possessing the highest speeds. The latter conclusion revives the idea of the Expanding Molecular Ring (EMR) at the boundary of the CMZ, proposed by \citet{kai72} and by \citet{sco72}. However, there are three essential differences: (1) The expanding gas viewed in H$_3^+$ does not form a ring but fills most of the volume of the CMZ; (2) the gas in the EMR is diffuse and thus much less dense than assumed previously; and (3) the gas is expanding purely radially, without the previously reported rotation. The first of these conclusions suggests that there is a one-to-one correspondence between radial velocity and radial location. As explained below for any CMZ sightline such a relation makes it possible to deduce the locations of continuum sources within the CMZ from observed velocity profiles of lines formed in the diffuse gas.
\setcounter{figure}{0}
\begin{figure}
\includegraphics[angle=-0,width=0.42\textwidth]{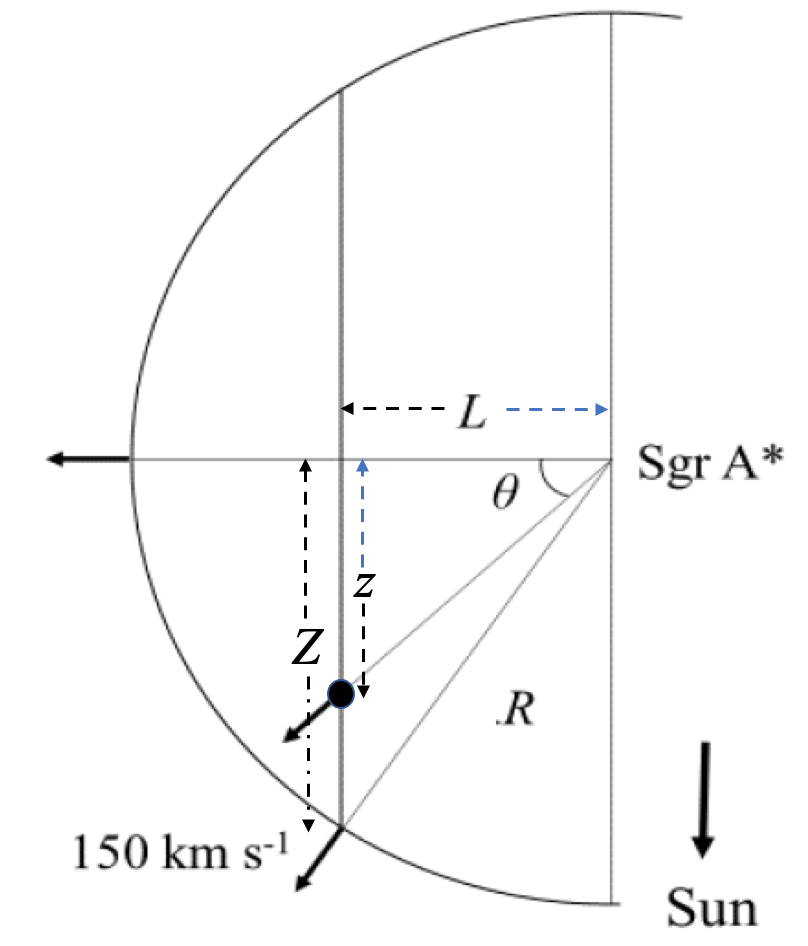}
\caption{Schematic face-on view of half of the CMZ. Diffuse gas filling the CMZ is expanding at $V$ = 150 km s$^{-1}$ at its outer circumferential surface, which is at radius $R$ = 150 pc from the central black hole, Sgr  A$^\ast$. The vertical line at distance $L$ from Sgr~A$^\ast$ is the line of sight, and has length Z from its midpoint to the front edge of the CMZ. For an assumed relation between expansion velocity and distance from Sgr~A$^\ast$, the radial velocity of the gas producing absorption at the location denoted by the dot, located a distance $z$ from the midpoint, can be calculated (see text for details).}
\end{figure}

In Figure 1, for a sightline displaced on the sky by $L$ from Sgr~A$^\ast$ the distance that its intersection with the circumferential surface of the CMZ is in front of Sgr~A$^\ast$ is $Z$ = $\sqrt{R^2 - L^2}$ and the radial velocity of the gas at the that location is  $V_Z$= $VZ/R$. The radial velocity $v_z$ at an interior point is unknown. To calculate it one would need to know the distribution of the gas velocities when the explosion occurred. (An estimate of the ejection velocity of the gas that is now at the front surface of the CMZ is given in the Appendix of Paper II.) Here we simply assume that the outward speed of the expanding gas along a radial ray at position $z$ is proportional to the distance from Sgr~A$^\ast$; that is,  
\begin{math}
v_{exp} = V\sqrt{L^2 + z^2}/R
\end{math}
and its radial velocity as viewed from the Sun is 

\begin{equation}
v_z = -V \frac{\sqrt{L^2 + z^2}}{R} \sin \theta = -V\frac{z}{R}.     
\end{equation}

This is a reasonable assumption in view of the results in Papers I and II. Using this relation, the depth of the position within the CMZ relative to Sgr~A$^\ast$ is determined from the observed radial velocity $v_{z}$ simply as $z$ = $-Rv_{z}/V$. How accurately this assumption represents the actual variation of the speed of the gas is uncertain but the observed roughly trapezoidal velocity profiles of the absorption lines of H$_3^+$ in diffuse gas (Paper II) qualitatively support it. 

\section{Determination of the Locations of Star $\iota$ and Sgr B2}

From Figure 1 it is apparent that for an absorption line arising in the diffuse gas observed in the spectrum of a continuum source located anywhere within the CMZ, the location of the source is defined by the more positive velocity edge of the absorption. In Section 3.1, using the approximate equation (1) we determine the location of Star $\iota$ in the CMZ from velocity profiles of the infrared absorption spectrum of  H$_3^+$, mainly observed by the Phoenix spectrometer \citep{hin98} on the Gemini South Telescope. Likewise, in Section 3.2 we determine the locations of the far-infrared continuum sources Sgr B2(M) and Sgr B2(N), two high mass star-forming dense molecular cores separated on the sky by $\sim$2 pc  \citep{san17} using far-infrared absorption spectra of H$_2$O$^+$, OH$^+$ and $^{13}$CH$^+$ observed by the Heterodyne Instrument for the Far-Infrared \citep[HIFI,][]{gra10} on the $Herschel$ Space Observatory. While H$_3^+$ resides both in dense($n$ $\gtrsim$ 5 $\times$ 10$^2$ cm$^{-3}$) and diffuse ($n$ $\lesssim$ 5 $\times$ 10$^2$ cm$^{-3}$) clouds \citep[see][for descriptions of their basic properties as they pertain to H$_3^+$]{mil20}, the latter three molecular ions can exist only in diffuse clouds, as explained in Section 3.2. 

The observed velocity extrema of the absorption lines of the above molecular species are listed in Table 1.  We also show and briefly discuss the velocity profiles of H$_3$O$^+$ and NH$_3$, which reside only in dense clouds. The results and their uncertainties are discussed in Section 4. 


\begin{figure}
\includegraphics[width=0.45\textwidth, angle=-0]{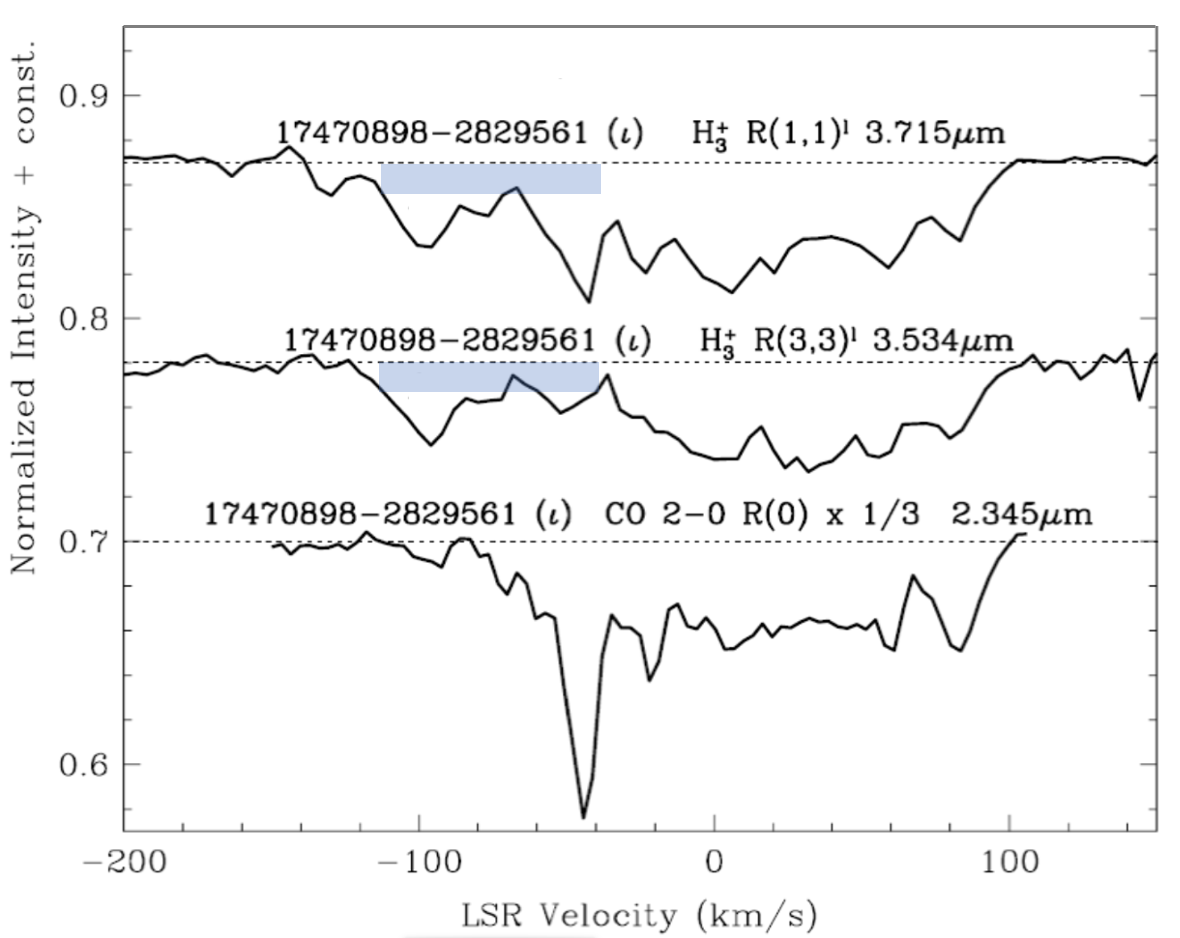}
\caption{ Velocity profiles of the H$_3^+$ $R$(1,1)$^l$ and $R$(3,3)$^l$ and the CO $v$ = 2$\leftarrow$0 $R$(0) absorption lines, reproduced from Figure 7 of Paper II. The resolution of the spectra is 6 km s$^{-1}$. The blue rectangles in the H$_3^+$ profiles are the velocity intervals where the absorption is due to radially expanding diffuse gas in the front half of the CMZ.}
\end{figure}

Figure 2, which reproduces a portion of Figure 7 of Paper II, contains the velocity profiles of the $R$(1,1)$^l$ and $R$(3,3)$^l$ lines of H$_3^+$ and the $v$ = 2$\leftarrow$0 $R$(0) line of CO. In Paper II absorption at velocities more negative than $-40$ km s$^{-1}$ was ascribed to H$_3^+$  in diffuse clouds, but the possibility was allowed that diffuse gas is also present at less negative velocities. At positive velocities the rough similarity of the two H$_3^+$ line profiles to the CO line profile seems to suggest that, because observable CO $v$ = 2$\leftarrow$0 absorption lines arises in dense gas (see Section 1.8 of Paper I), most of the H$_3^+$ absorption toward Star $\iota$ also arises in that environment.  However this conclusion is incorrect, as explained below.

Although H$_3^+$ resides in both dense and diffuse clouds \citep[see, e.g.,][]{oka13, mil20}, in the CMZ one can distinguish between the two environments by comparing the profiles of three vibration-rotation lines, arising from the ($J$,$K$) = (1,1), (2,2), and (3,3) levels of the ground vibrational state (only two of which are shown in Figure 2). This is because although the (1,1) and (3,3) levels are populated both in dense clouds and in diffuse clouds (the latter level only when the diffuse gas is sufficiently warm, which is the case in the CMZ; see Paper I), the (2,2) level is populated only in clouds that are both warm and dense. Because H$_3^+$ is nonpolar due to its equilateral triangular structure, ordinary rotational transitions are forbidden. However, symmetry-breaking forbidden transitions are allowed \citep{oka71, wat71, pan86}. Thus the pure rotational spontaneous transition (2,2)$\rightarrow$(1,1) can occur; it has a lifetime of $\sim$27 days \citep{nea96}, for which the corresponding critical density is approximately 200 cm$^{-3}$. Thus, the $R$(2,2)$^l$ line can only be observed when $n$ is higher than 200 cm$^{-3}$ and therefore can test if warm gas is dense or diffuse.

\begin{deluxetable}{llcc}
\footnotesize
\tablecaption{Velocity Ranges of Absorbing Diffuse Gas \label{t1}}
\tablehead{
\colhead{Source} &
\colhead{Molecular Transition} &
\colhead{v$_{LSR}$(min)} &
\colhead{v$_{LSR}$(max)} \\
\colhead{} &
\colhead{} &
\colhead{km s$^{-1}$} &
\colhead{km s$^{-1}$}
}
\startdata
Star $\iota$ & H$_3^+$ $v$ = 1$\leftarrow$0 $R$(1,1)$^l$ & $-115$ & $+90$ \\ \\
Sgr B2(M) & o-H$_2^+$ 1$_{11}$$\leftarrow$0$_{00}$ $J$ = 2.5$\leftarrow$1.5 & $-148$ & $+90$ \\
Sgr B2(M) & OH$^+$ $N$ = 1$\leftarrow$0, $J$ = 0$\leftarrow$1 & $-132$ & $+88$ \\
Sgr B2(M) & OH$^+$ $N$ = 1$\leftarrow$0, $J$ = 2$\leftarrow$1 & $-146$ & $+92$ \\
Sgr B2(M) & OH$^+$ $N$ = 1$\leftarrow$0, $J$ = 1$\leftarrow$1 & $-143$ & $+88$ \\
Sgr B2(M) & Mean & $-142$ & $+89$ \\ \\
SgrB2(N) & OH$^+$ $N$ = 1$\leftarrow$0, $J$ = 0$\leftarrow$1 & $-132$ & $+94$ \\
Sgr B2(N) & OH$^+$ $N$ = 1$\leftarrow$0, $J$ = 2$\leftarrow$1 & $-135$ & $+99$ \\
Sgr B2(N) & OH$^+$ $N$ = 1$\leftarrow$0, $J$ = 1$\leftarrow$1 & $-132$ & $+97$ \\
Sgr B2(N) & $^{13}$CH$^+$ $N$ = 1$\leftarrow$0 & $-136$ & $+96$ \\
Sgr B2(N) & Mean & $-134$ & $+96$
\enddata
\end{deluxetable}

\begin{figure}
\includegraphics[width=0.45\textwidth, angle=-0]{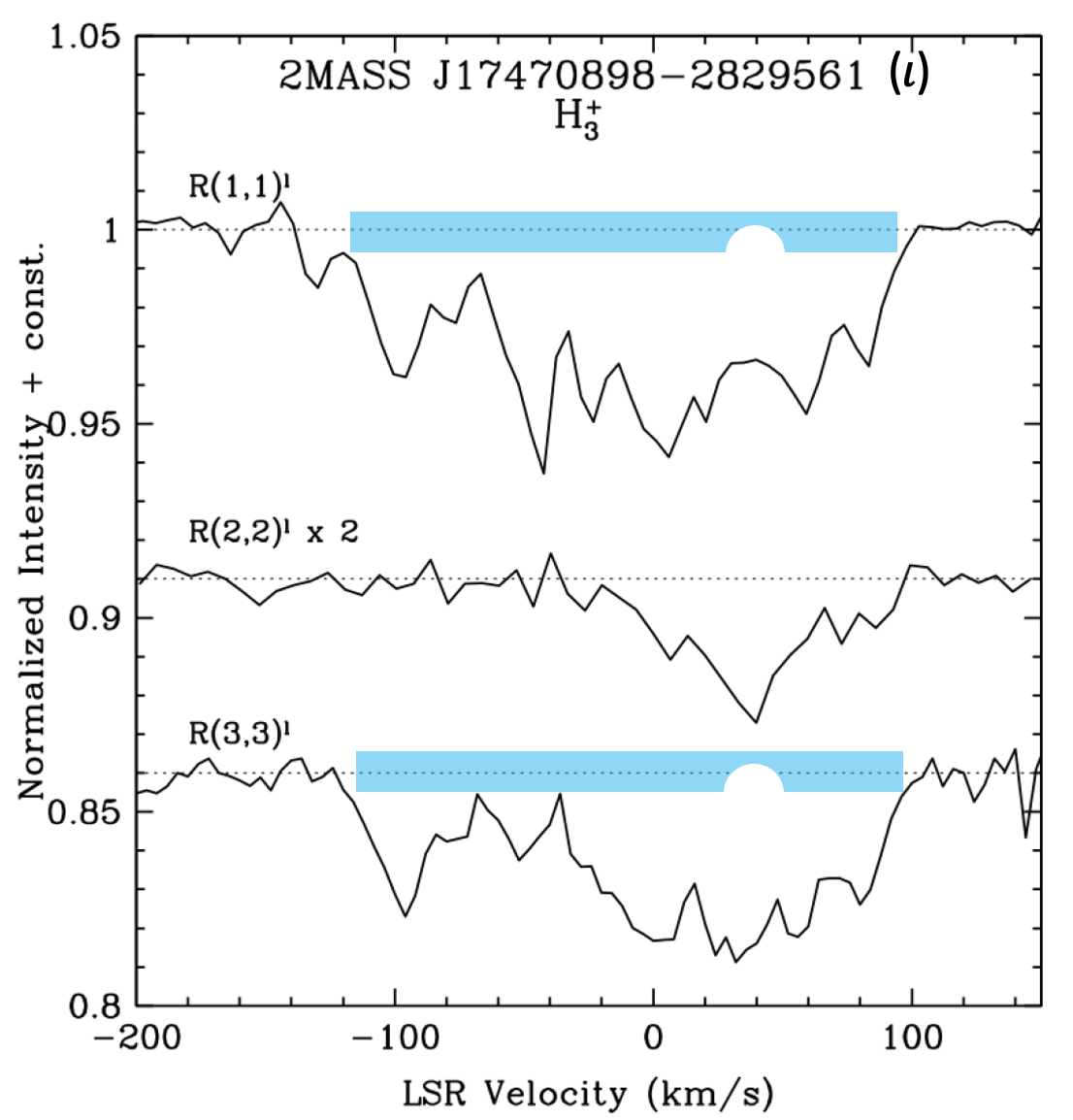}
\caption{Observed velocity profiles of the $R$(1,1)$^l$, $R$(2,2)$^l$, and $R$(3,3)$^l$ absorption lines of H$_3^+$ toward Star $\iota$, from \citet{geb21}. The resolution is 6 km s$^{-1}$. Note that unlike the other two lines the $R$(2,2)$^l$ profile has a maximum absorption at $+40$ km s$^{-1}$ and is significantly weaker or absent elsewhere; in contrast the other lines  have roughly constant depths at positive velocities from 0 to 80 km s$^{-1}$. This difference indicates that diffuse gas (blue shaded regions) is present over virtually the entire positive velocity range as well as at negative velocities (see text). The narrowing of the shaded region near +40 km s$^{-1}$ indicates a large contribution by H$_3^+$ in dense gas.}

\end{figure}

\begin{figure}
\includegraphics[width=0.42\textwidth, angle=-0]{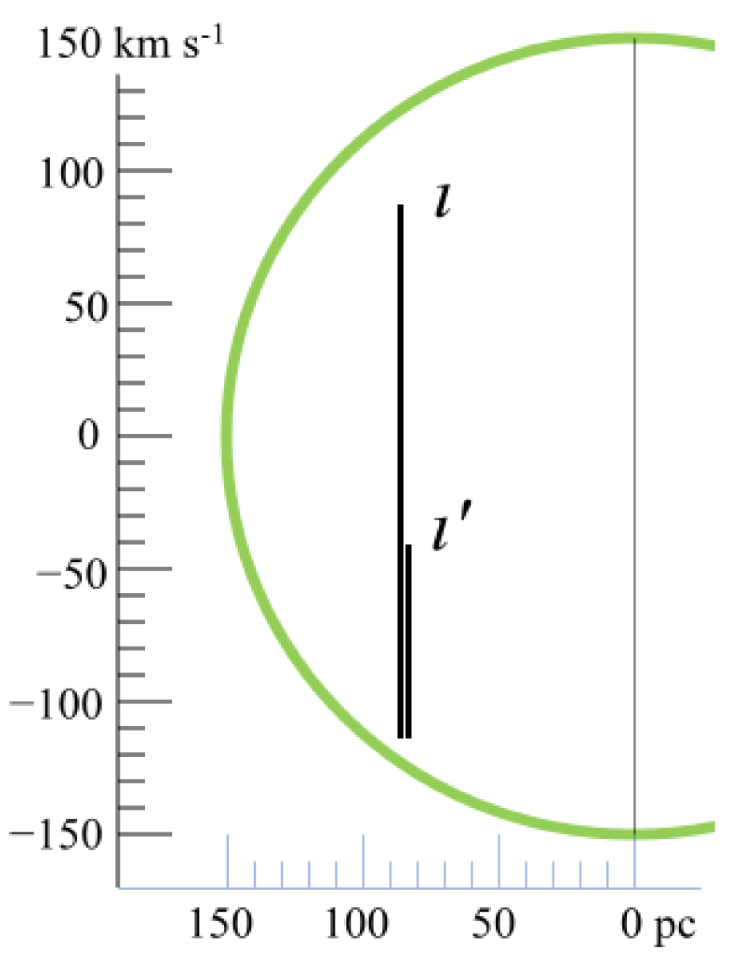}
\caption{The revised ($l$, $v$) diagram for H$_3^+$ in diffuse gas toward Star $\iota$, extending from $-115$ to $+90$ km s$^{-1}$, indicating that the star is 90 pc more distant than Sgr~A$^\ast$ (see text). The shorter bar denoted as $\iota$’ (shifted horizontally for clarity) corresponds to the rectangles in Figures 2 and 11 of Paper II, which are corrected in this paper.
 }

\end{figure}

The velocity profiles toward Star $\iota$ of lines from all three of the above levels are shown in Figure 3. Based on the $R$(2,2)$^l$ profile, H$_3^+$ in dense clouds makes a significant contribution to the absorption near $+40$ km s$^{-1}$ but is much less prominent at other positive velocities, although the weak absorption indicates that dense molecular gas extends to a velocity of $+90$ km s$^{-1}$. At the velocity of maximum $R$(2,2)$^{l}$ absorption, i.e. at  +40 km s$^{-1}$, the observed depths of the $R$(1,1)$^{l}$, $R$(2,2)$^{l}$, and $R$(3,3)$^{l}$ lines are 3.61\%, 1.34\%, and 4.39\%, respectively. Using the absorption strengths of these lines (see Table 1 of Paper I), the level  populations of (1,1), (2,2) and (3,3) at that velocity are in the ratios 1.00 : 0.30 : 0.83. Away from +40 km s$^{-1}$ the strength of the $R$(2,2)$^{l}$ absorption rapidly decreases. In contrast both the $R$(1,1)$^l$ and $R$(3,3)$^l$ profiles are of roughly constant depth from 0 to 80 km s$^{-1}$. This implies that in addition to dense gas, warm diffuse gas is present across virtually the entire positive velocity range of the absorption. Since the contributions of dense clouds and diffuse clouds to the $R$(1,1)$^l$ and $R$(3,3)$^l$ absorption profiles cannot be separated, one cannot easily determine the temperatures and densities of H$_3^+$ in the dense gas. For the purposes of this paper, however, it suffices to note that with the possible exception of radial velocities near $+40$ km s$^{-1}$ a large fraction of the H$_3^+$ resides in diffuse gas. In Figure 3 we therefore indicate the H$_3^+$ in diffuse clouds as extending from $-115$ to $+90$ km s$^{-1}$. Inserting $+90$ km s$^{-1}$  into equation (1) yields $z$ = $-90$ pc; in other words, Star $\iota$ lies 90 pc behind Sgr~A$^\ast$. The revised ($l$, $v$) diagram for Star $\iota$, corrected from that of Paper II, is shown in Figure 4.

\subsection{Locations of Sgr B2(M) and Sgr B2(N)}

  Unlike H$_3^+$ which resides both in dense and diffuse clouds, the cations H$_2$O$^+$, OH$^+$, and CH$^+$ reside only in diffuse clouds. This is because H$_2$ destroys them through efficient hydrogen abstraction reactions: H$_2$O$^+$ +  H$_2$ $\rightarrow$ H$_3$O$^+$ + H; OH$^+$ + H$_2$ $\rightarrow$ H$_2$O$^+$ + H; and CH$^+$ + H$_2$ $\rightarrow$ CH$_2^+$ + H, respectively with high Langevin rate constants, 0.83$\times$10$^{-9}$, 1.00$\times$10$^{-9}$, and 1.20$\times$10$^{-9}$ cm$^{3}$ s$^{-1}$, respectively \citep{ani86}. Thus, their observed velocity profiles correspond purely to velocities of diffuse gas. A complication for the H$_2$O$^+$ and OH$^+$ absorption lines is the presence of hyperfine structure which complicates their profiles.

\begin{figure}
\includegraphics[width=0.48\textwidth, angle=-0]{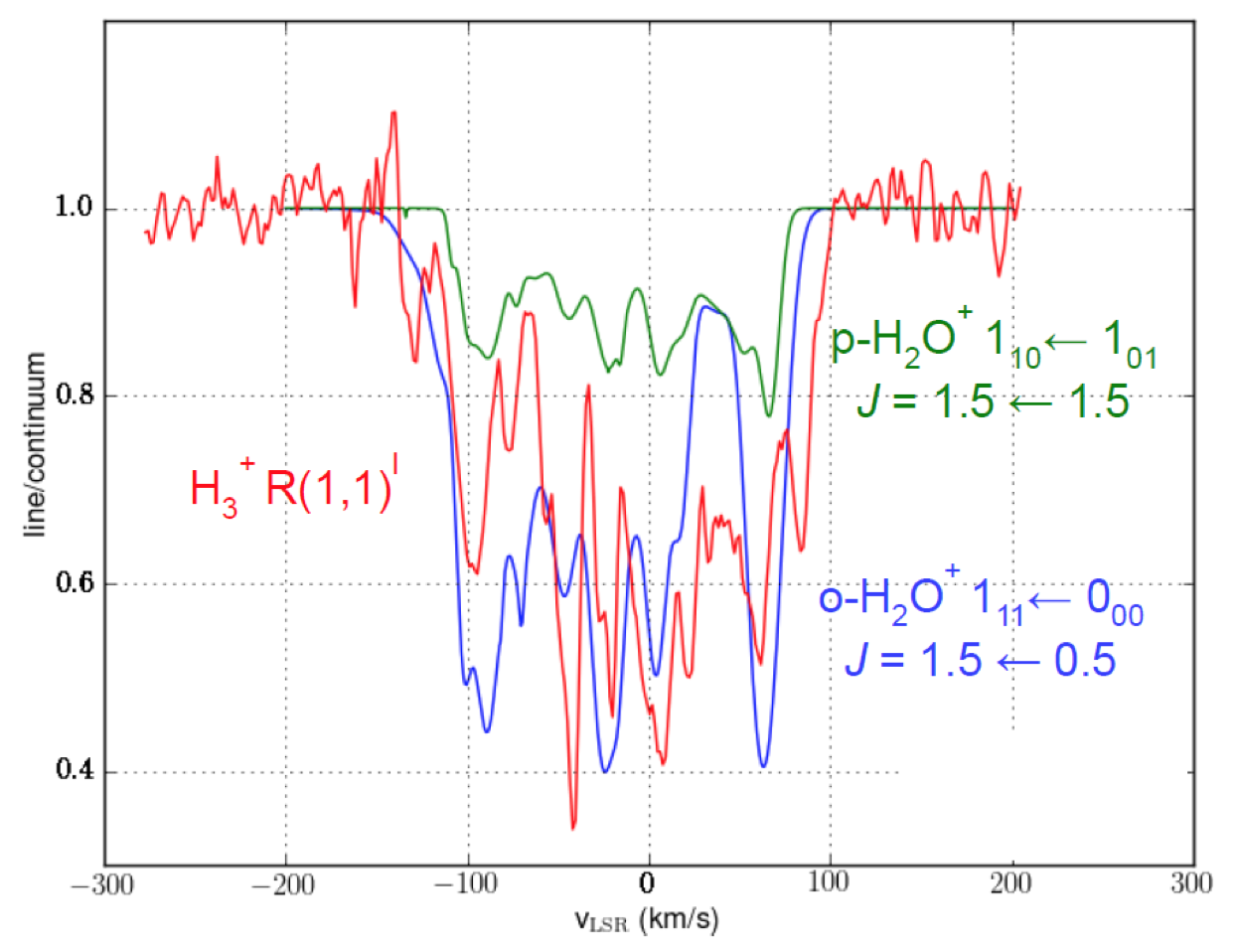}
\caption{The velocity profiles of the $R$(1,1)$^l$ line toward Star $\iota$ \citep[][red]{geb10} and two (ortho and para) rotational lines of H$_2$O$^+$ (blue and green, respectively) toward Sgr B2(M) \citep{sch10}. The H$_3^+$ profile is multiplied by a factor of 10 for easier comparison.}
\end{figure}

\begin{figure}
\includegraphics[width=0.48\textwidth, angle=-0]{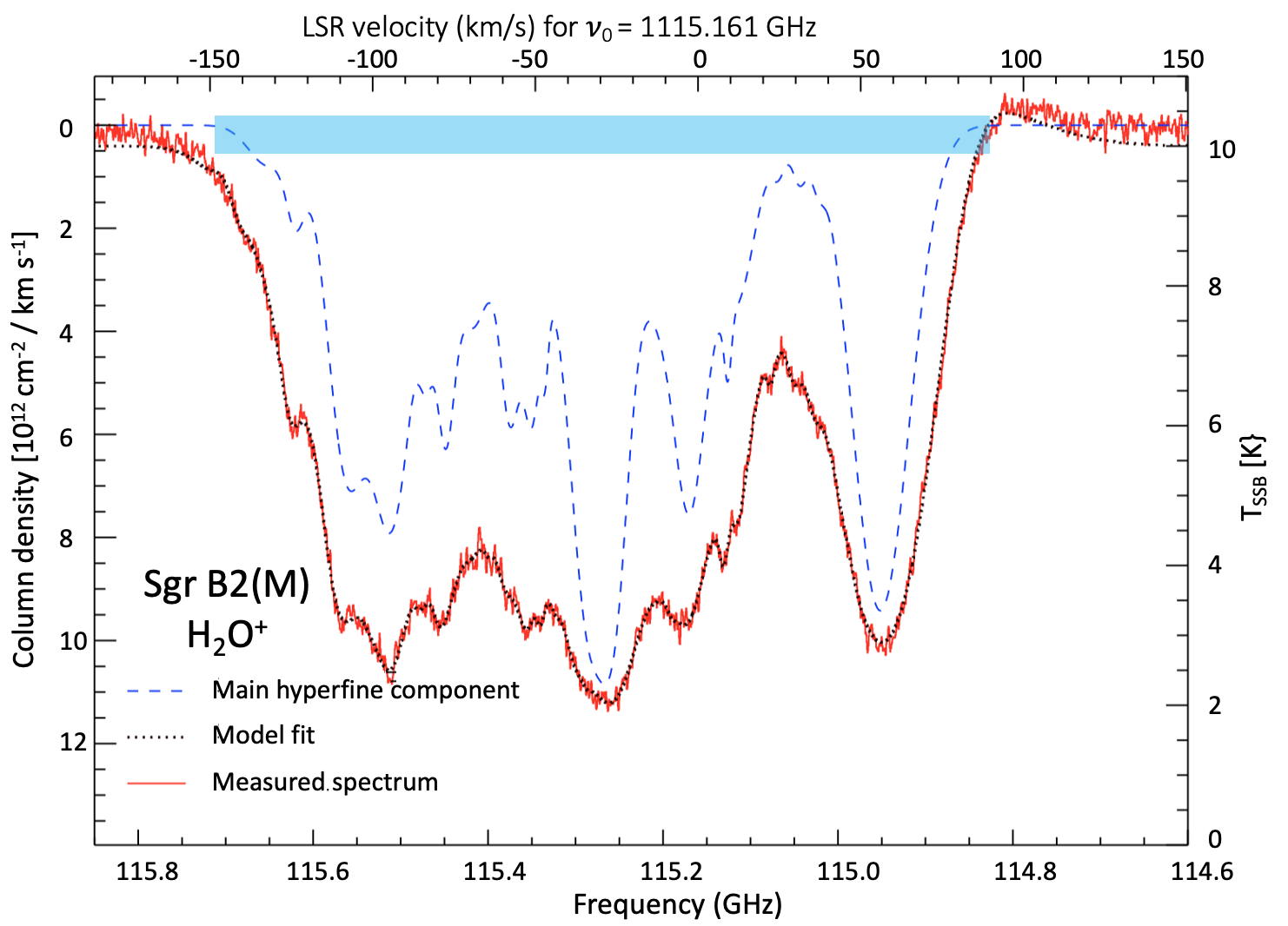}
\caption{The observed profile of the ortho H$_2$O$^+$ 1$_{11}$ $\leftarrow$ 0$_{00}$ $J$ = 1.5 $\leftarrow$ 0.5 absorption line 
toward Sgr B2(M), in which five hyperfine components overlap, and the extracted velocity profile of the main component, $F$ = 2.5$\leftarrow$1.5 (blue dashed line). The rectangle indicates the velocity range of the diffuse gas. 
Adapted from \citet{oss10}.
}
\end{figure}

\subsubsection{Comparison of Velocity Profiles of H$_3^+$ toward Star $\iota$ and H$_2$O$^+$ toward Sgr B2(M)}

The absorption profile of the $R$(1,1)$^l$ vibration-rotation line of H$_3^+$ toward Star $\iota$ is compared to the ortho-H$_2$O$^+$ 1$_{11}$ $\leftarrow$ 0$_{00}$ $J$ = 1.5$\leftarrow$0.5 and para-H$_2$O$^+$ 1$_{10}$ $\leftarrow$ 0$_{01}$ $J$ = 1.5$\leftarrow$0.5 rotational transitions toward Sgr B2(M) \citep{sch10} in Figure 5. Since the permanent dipole moment of H$_2$O$^+$ \citep[2.37 D,][]{wu04} is $\sim$15 times greater than the transition dipole moment of H$_3^+$ \citep[0.158 D,][]{car76}, for equal abundances the H$_2$O$^+$ lines would be $\sim$200 times stronger than the H$_3^+$ line. The observed H$_3^+$ absorption depth toward Star $\iota$ is about one-tenth of the depth of the ortho-H$_2$O$^+$ absorption toward Sgr B2(M), indicating that the column density of ortho-H$_2$O$^+$  is about one-twentieth of the column density of H$_3^+$.
 
It is apparent from Figure 5 that the velocity profiles of H$_3^+$ toward Star $\iota$ and those of ortho- and para-H$_2$O$^+$  toward Sgr B2(M) are similar in their velocity ranges and in some of their detailed structures. However, both the ortho- and para-H$_2$O$^+$ lines contain hyperfine structure, which is responsible for the differences in their profiles. We use the velocity profile of the strongest ortho hyperfine component derived by \citet[][Figure 4]{oss10}, which is reproduced in Figure 6, for deducing more precisely the velocity extent of the diffuse gas in which the H$_2$O$^+$ is located.

The velocity resolution of HIFI is $\ll$ 1 km s$^{-1}$ and the signal-to-noise ratios of the H$_2$O$^+$  and OH$^+$ spectra are very high. Therefore, unlike the profiles of the H$_3^+$ lines, whose full extents are uncertain by $\sim$10 km s$^{-1}$, the actual extents of the far-infrared absorption lines are essentially the observed extents. From the main hyperfine component in Figure 6 we observe a full velocity range of $-148$ to $+90$ km s$^{-1}$. From Equation (1) the positive velocity limit places Sgr B2(M) behind  Sgr~A$^\ast$ by $\sim$90 pc. Thus, to within the uncertainties Star $\iota$ and Sgr B2(M), although separated on the sky by 17 pc, are located the same distance behind Sgr~A$^\ast$. 

\begin{figure}
\includegraphics[width=0.48\textwidth, angle=0]{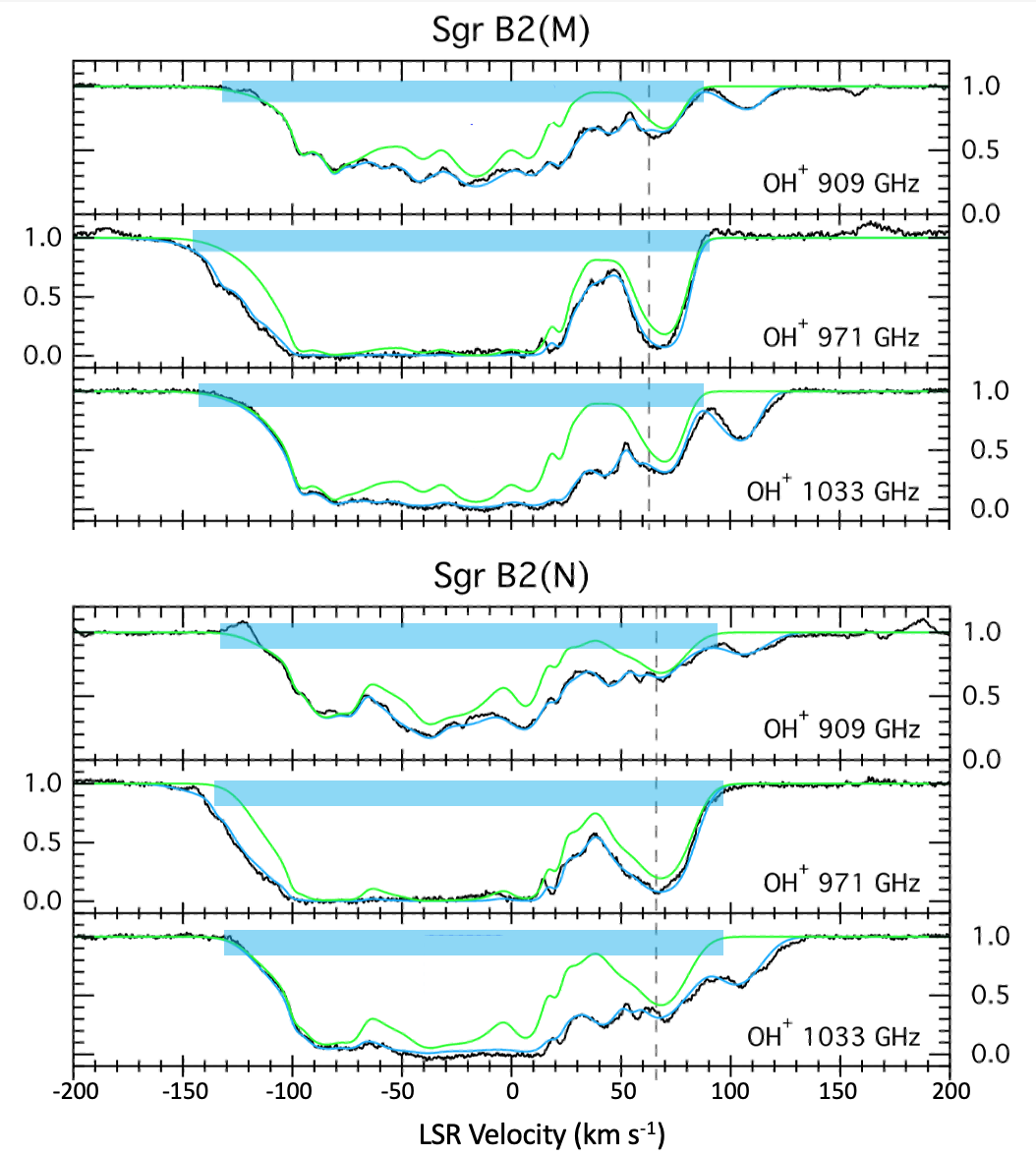}
\caption{Normalized velocity profiles of the OH$^+$ $N$ = 1 $\leftarrow$ 0 fine structure components $J$ = 0 $\leftarrow$ 1 (top), 2 $\leftarrow$ 1 
(middle), and 1  $\leftarrow$ 1 (bottom) lines toward Sgr B2(M) and Sgr B2(N) \citep[from][]{ind15}. The green traces represent the strongest hyperfine components. The rectangles are the velocity ranges of the absorbing gas of these components.
}
\end{figure}

\subsubsection{Velocity Profiles of OH$^+$ and CH$^{+}$ and the Locations of Sgr B2(M) and Sgr B2(N)}

The velocity profiles of three OH$^+$ lines observed by \citet{ind15} toward Sgr B2(M) and Sgr B2(N) are shown in Figure 7. Eyeball estimates of the velocity range for each fine structure component are given in Table 1. The positive velocity edges of the absorptions for Sgr B2(M) for these lines as well as the H$_2$O$^+$ line presented in the previous section agree with one another to a few kilometers per second (see Table 1); their average value is 89 km s$^{-1}$, which differs from the value for H$_2$O$^+$ by only 1 km s$^{-1}$. The positive velocity edges for Sgr B2(N) average to 97 km s$^{-1}$, slightly higher than Sgr B2(M). This indicates that Sgr B2(N) lies slightly behind Sgr B2(M) and is further to the rear of the CMZ than either Sgr B2(M) or Star $\iota$.

\begin{figure}
\includegraphics[width=0.42\textwidth, angle=-0]{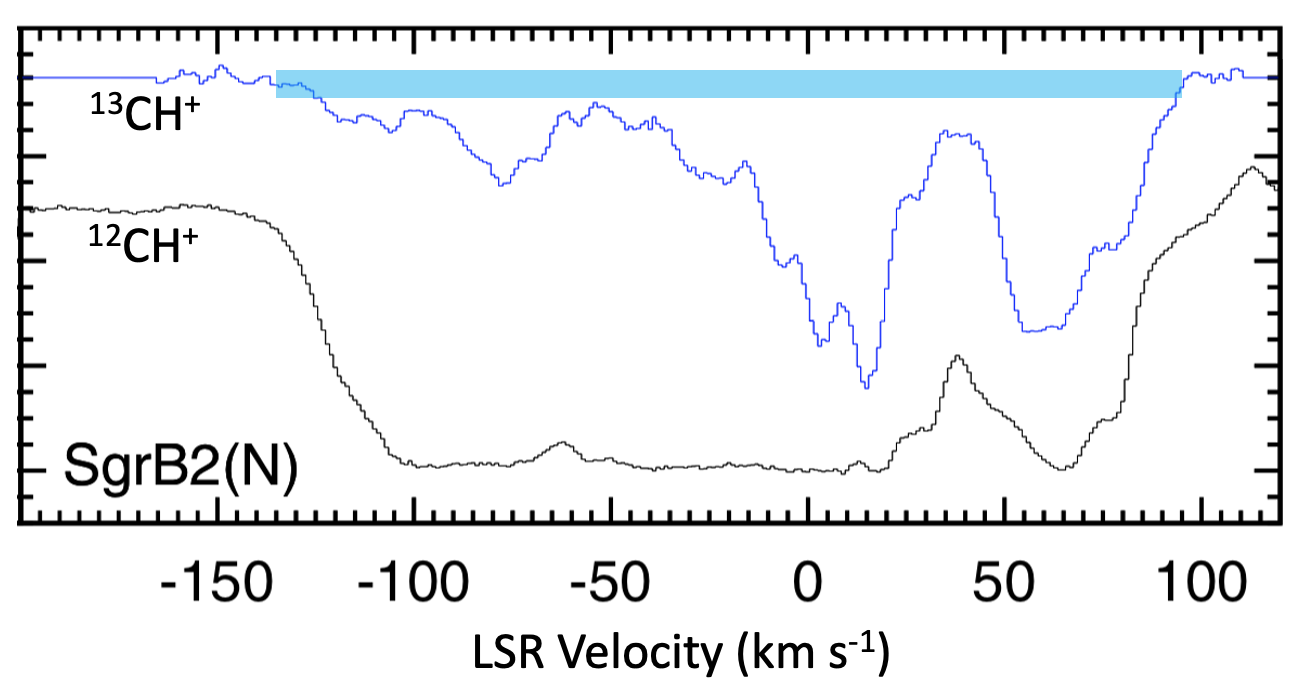}
\caption{Velocity profiles of the $^{13}$CH$^+$ and $^{12}$CH$^+$ $N$ = 1$\leftarrow$0 absorption lines observed toward Sgr B2(N) reported by \citet{god12}. Because of the extreme saturation of the $^{12}$CH$^+$ line we use the $^{13}$CH$^+$ line to estimate the velocity extrema.
}
\end{figure}

Unlike the triplet cation OH$^+$, which has $^3$$\Sigma^-$ symmetry, $^{12}$CH$^+$ is of symmetry  $^1$$\Sigma$ and its spectrum does not have hyperfine structure \citep{ama10a}. That is not the case for $^{13}$CH$^+$; its splitting, 1.63 MHz, is anomalously large, as explained by \citet{ama10b}. In terms of velocity, however, that splitting corresponds to 0.59 km s$^{-1}$ and thus is unimportant here.

The observed velocity profiles of the $N$ = 1$\leftarrow$0 transition of $^{12}$CH$^+$ and $^{13}$CH$^+$ toward Sgr B2(N) reported by \citet{god12} are shown in Figure 8. We estimate the positive edge of the profile to be 96 km s$^{-1}$, consistent with the values found from OH$^+$ for this source.

\begin{figure}
\includegraphics[width=0.45\textwidth, angle=-0]{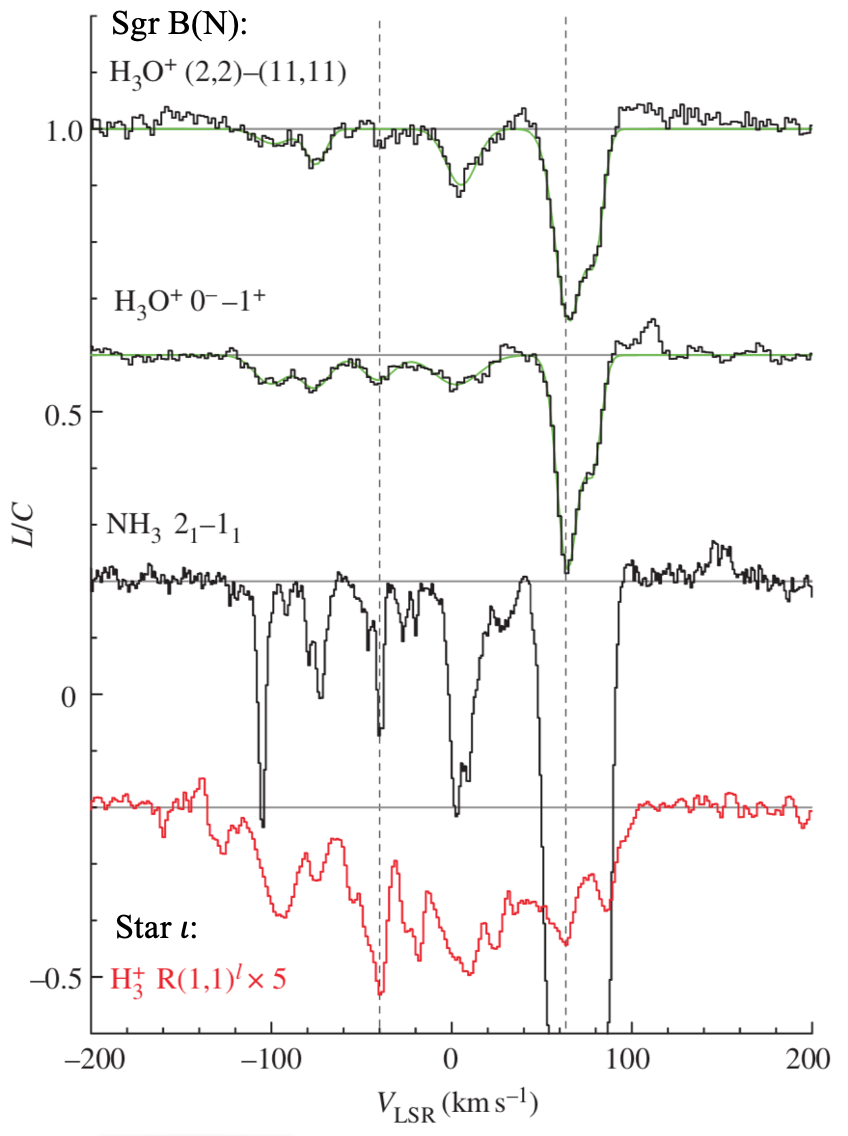}
\caption{Velocity profiles of a uniformly weighted average of the H$_3$O$^+$ ($J$,$K$) = (2,2)$-$(11,11) inversion transitions (top trace), the H$_3$O$^+$ (0,0)$^-$ $\leftarrow$ (1,0)$^+$ transition (second trace), and the NH$_3$ (2,1)$^+$ $\leftarrow$ (1,1)$^-$ transition (third trace), all toward Sgr B2(N), and the H$_3^+$ $R$(1,1)$^l$ line toward Star $\iota$. From \citet[][Figure 4]{lis12}. 
}
\end{figure}

\subsection{Comparisons of NH$_3$ and H$_3$O$^+$ Absorption Line Profiles toward Sgr B2 with H$_3^+$ Absorption Toward Star $\iota$.}

Although in this paper the dense clouds in Sgr B2 are not used to estimate the location of Sgr B2, it is of interest from a chemical perspective to compare the spectrum of H$_3^+$ toward Star $\iota$ to spectra of NH$_3$ and H$_3$O$^+$ obtained by \citet{lis12}. Unlike OH$^+$ and H$_2$O$^+$ the fully protonated H$_3$O$^+$ does not react with H$_2$ and therefore is stable in dense clouds. Also unlike paramagnetic OH$^+$ and H$_2$O$^+$, H$_3$O$^+$, like the isoelectronic molecule NH$_3$, has no fine structure and only minute hyperfine structure. 

In many ways H$_3$O$^+$  is similar to NH$_3$ both chemically and spectroscopically, but there are two notable differences. (1) The abundance of H$_3$O$^+$ is less than that of NH$_3$ because it is destroyed by dissociative electron recombination (although the effect of this on its abundance is much less in dense clouds than diffuse clouds) and because the proton affinity of H$_2$O (7.22 eV) is less than those of some molecules that are abundant in dense clouds such as NH$_3$ (8.85 eV) and HCN (7.43 eV). See, e.g., Table 1 of \citet{oka13}. Thus, NH$_3$ destroys H$_3$O$^+$ through the proton hop reaction H$_3$O$^+$ +  NH$_3$ $\rightarrow$ H$_2$O + NH$_4^+$, which has a high exothermicity,1.42 eV, and a large Langevin rate constant, $\sim$2.4 $\times$ 10$^{-9}$  cm$^3$ s$^{-1}$ \citep{ani86}. (2) The central charge of  H$_3$O$^+$ attracts its three hydrogen atoms, resulting in the OH bonds being stronger than the  NH bonds, making the bonds more like $sp2$ than $sp3$, and causing the H$_3$O pyramid to be shallower than the NH$_3$ pyramid. This makes the frequency of the H$_3$O$^+$ inversion, 55.346 cm$^{-1}$ \citep{liu85} nearly 70 times higher than that of NH$_3$, 0.793 cm$^{-1}$.

Figure 9 shows velocity profiles of lines of both species toward Sgr B2(N) as well as H$_3^+$ toward Star $\iota$. It can be seen that the absorption components in the H$_3$O$^+$ and NH$_3$ lines are at the same velocities, indicating that those species reside in the same dense clouds although absorption in the latter species is more than 10 times stronger than in the former. Their profiles bear only vague resemblances to the profile of the H$_3^+$ line. This is not surprising in that (1) the H$_3^+$ absorption line contains contributions both from dense gas and diffuse gas and (2) its sightline is displaced by $\sim$17 pc. They also are vastly different from the CO absorption profile toward Star $\iota$ shown in Figure 2. This indicates that although the H$_3$O$^+$ and NH$_3$  seen toward Sgr B2 and the CO seen toward Star $\iota$ reside in dense clouds the clouds on their sightlines must have considerably different physical properties. Indeed it is well known that Sgr B2(M) and Sgr B2(N) contain dense cores \citep[e.g., see][]{san17}, whereas none are located on the sightline to Star $\iota$ \citep{geb21}.

\begin{figure}
\includegraphics[width=0.48\textwidth, angle=-0]{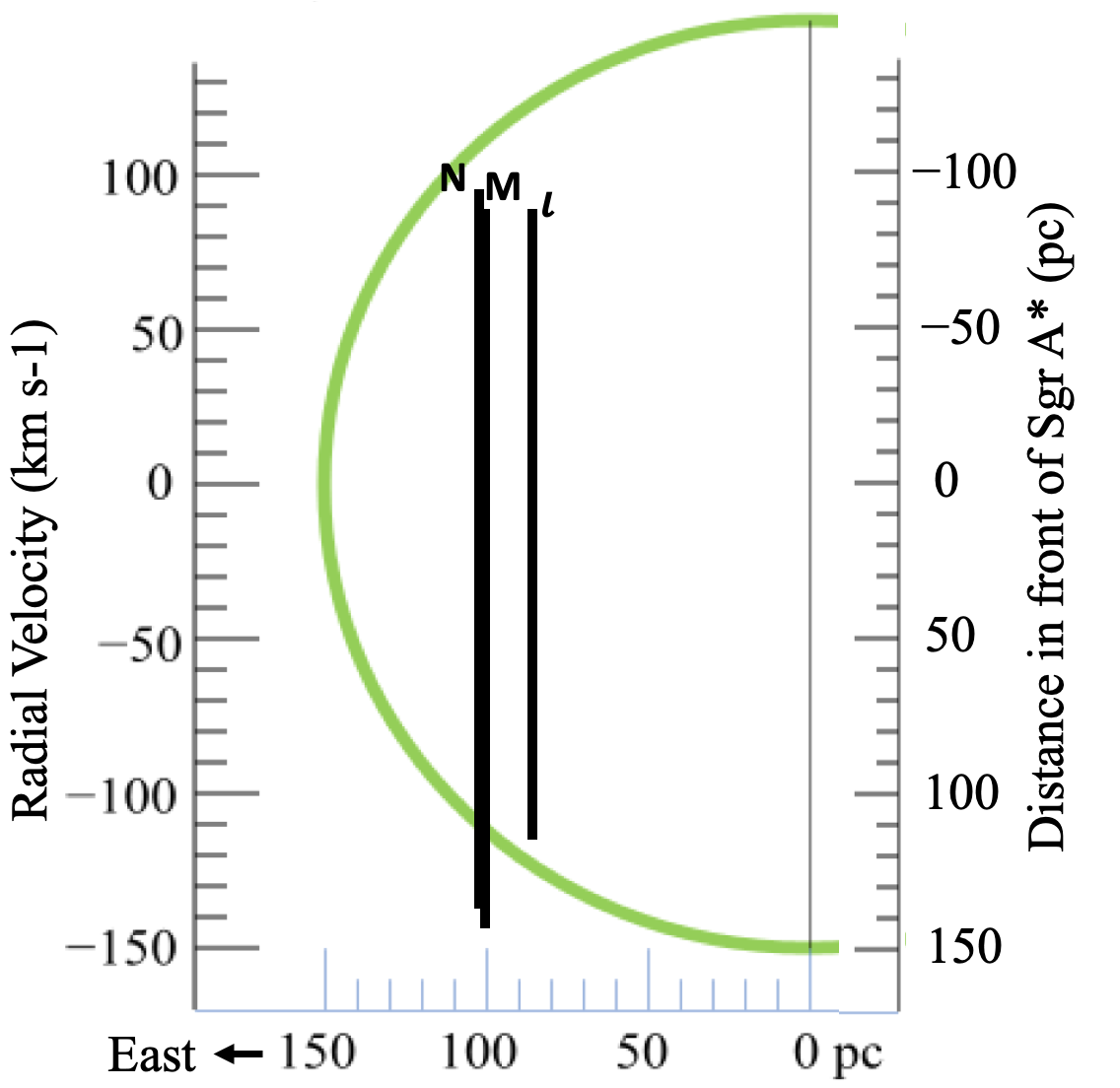}
\caption{The ($l$, $v$) diagram of the diffuse gas columns toward Star $\iota$ and the hot molecular cores Sgr B2(M) and Sgr B2(N). The vertical axis on the right is a translation of velocity into depth within the CMZ relative to Sgr~A$^\ast$ using Equation 1.
} 
\end{figure}

\section{Discussion}

\subsection{The Locations of Star $\iota$ and Sgr B2 in the CMZ}

Figure 10 summarizes the main result of this work, which is based on the simple model in Section 2 for the velocity field of the warm diffuse gas in the CMZ, previously shown to be undergoing radial expansion from an origin at or near Sgr~A$^\ast$ (Paper II). Star $\iota$ and Sgr B2 are both approximately 90 pc more distant than Sgr~A$^\ast$. Such large distances behind Sgr~A$^\ast$ puts them near the rear of the CMZ. Based on the model and to within the measurement uncertainties Sgr B2(N) lies slightly behind Sgr B2(M). The above conclusions are based on the positive velocity extremum of the observed absorption by the H$_3^+$ $R$(1,1)$^l$ line toward Star $\iota$ and the average of the positive velocity extrema of the observed absorption lines of three molecular ions, H$_2$O$^+$, OH$^+$, and $^{13}$CH$^+$ toward the Sgr B2 far-infrared continuum sources.

\subsection{Uncertainties}

The uncertainty in our estimate of the positive velocity extremum of the H$_3^+$ line toward Star $\iota$ is approximately 10 km s$^{-1}$. For the far-infrared lines observed which were observed toward Sgr B2 at much higher spectral resolution and much higher signal-to-noise ratios, the uncertainties in the estimated extrema are only a few kilometers per second. It is noteworthy that for all of the far-infrared lines toward each of the Sgr B2(M) and Sgr B2(N), the estimated positive velocity extrema of each of the lines agree to within the above uncertainty. 

An additional source of uncertainty exists in the lines with significant hyperfine structure, which requires isolating a single hyperfine component to obtain accurate velocity extrema. One might expect that extrema  of the $^{13}$CH$^+$ profile, where the hyperfine structure is negligible (see Section 3.1.3.) and from OH$^+$, which has simpler hyperfine structure than H$_2$O$^+$, should be more reliable than the value derived from H$_2$O$^+$. Note, however, that H$_2$O$^+$ was only observed toward Sgr B2(M) and the velocity maximum  derived from it is fully consistent with those derived from the lines of OH$^+$.

One might question if the use of extreme positive edges of the far-infrared velocity profiles is the most appropriate way to estimate the locations of background source. In the modeled spectra of the hyperfine components of H$_2$O$^+$ and OH$^+$ only a miniscule fraction of the gas in the column has that velocity. It is conceivable that the motion of that gas is the result in small part of other processes besides the global radial expansion, such as interactions of the expanding gas with winds from some of the massive stars in Sgr B2. In addition, although on coarse scales the mass distribution in the inner few hundred parsecs of the Galaxy is isotropic \citep{sof13}, Sgr B2 and Sgr B1 are probably local peaks in mass density and as such could affect the velocity field of the expansion in their vicinities.  Our crude estimate of this effect is less than 10 km s$^{-1}$.

Finally, the derived distances to these three objects are dependent on the expansion model for the diffuse gas. We have chosen the simplest of these models, radial velocity linearly proportional to distance from Sgr~A$^\ast$, which assumes a single expulsion event centered on Sgr~A$^\ast$,  and a constant deceleration of the expanding gas distant from Sgr~A$^\ast$,  by more than 30 pc (see Paper II, Section 6 and its Appendix). Actual diffuse gas motions within the CMZ differing from this model by a few tens of km s$^{-1}$ would not be inconsistent with the observed data in Paper II, but would not change our conclusion.

The effects of most of the above could shift the positive velocity absorption edge in either direction and it is difficult to quantify the possible magnitudes of some of them. However, we believe based on the observed organized radial expansion of the diffuse gas in the CMZ with maximum observed speeds at the edge of the CMZ and the high positive velocity extrema of the absorption lines in that gas toward Star $\iota$, Sgr B(M), and Sgr B2(N), that the conclusion that these three objects lie far behind Sgr~A$^\ast$, is inescapable.

The much larger differences in the negative absorption extrema toward these three sources have no bearing on the results of this paper and also are not surprising. The edge of the CMZ where the expanding gas encounters the Galactic interstellar medium, may not be circular or have a sharp boundary and the ejection velocities may be somewhat different in different directions. In addition there is nothing to sharply truncate the absorption on that edge, as there is at positive velocities in the form of the continuum sources Star $\iota$, Sgr B2(M) and Sgr B2(N), which are embedded in the CMZ.

\subsection{Chemistry of Diffuse CMZ Gas toward Sgr B}

The continuous velocity profiles of the $R$(1,1)$^l$ and $R$(3,3)$^l$ absorptions of H$_3^+$ in Figure 3, from high negative to high positive values, indicate that H$_3^+$  in diffuse clouds exist largely uniformly within the CMZ on the sightline toward Star $\iota$. As pointed out in Paper II and in \citet{geb21} the deep and narrow absorption in the the $R$(1,1)$^l$  spectrum at  $-43$  km s$^{-1}$ arises in dense clouds in the 3 kpc spiral arm \citep{rou60}. Additional less prominent  narrow absorption by H$_3^+$ due to dense gas in the 4.5 kpc spiral arm \citep{men70} and the local spiral arm near 25 and 0 km s$^{-1}$, respectively, are also apparent in the $R$(1,1)$^l$ spectrum. These absorption components are also observed in many of the stars more centrally located in the CMZ \citep[see, e.g., Figure 1 of][and Paper II]{oka05}. The small shifts in radial velocities of these components toward stars within the CMZ are due to well-known velocity gradients along the foreground arms. 

The above three components are absent in the velocity profiles of the $R$(3,3)$^l$ line because the (3,3) level is not populated in low temperature gas. Apart from the spiral arm absorption features the shapes of the $R$(1,1)$^l$ and $R$(3,3)$^l$ absorptions profiles are similar, as is the case along the more central CMZ sightlines such as the one toward the brightest GCS 3-2 star in the Quintuplet Cluster \citep[see Figure 1 of][]{oka05}. The continuous absorption by H$_3^+$ over wide velocity ranges on many sightlines implies that the warm diffuse gas containing H$_3^+$ and therefore also containing H$_2$  are nearly ubiquitous in the CMZ (Paper I).

The velocity profiles of the H$_2$O$^+$, OH$^+$, and $^{13}$CH$^+$ absorption lines (Figures 6, 7, and 8, respectively), which also arise in the diffuse gas, indicate, however, that these molecular ions are not distributed as uniformly throughout the CMZ as H$_3^+$. Especially remarkable is the near complete absence of all three molecules near $v_{LSR}$ = $+40$ km s$^{-1}$. Whether this is due to chemistry, a true absence of diffuse gas at that velocity, or something else remains to be seen. We note that the velocity profiles of H$_3^+$ (Figure 3) indicate a possible paucity of diffuse gas at this velocity, on a sightline 17 pc distant from sightlines to the Sgr B2 sources. Perhaps this paucity extends across Sgr B2.

\section{Conclusion}

Assuming a linear relation between radial velocity and radial location of the diffuse gas in the CMZ, which is based on the data and conclusion in Paper  II of this series, we have used the maximum absorption velocities in spectra of H$_3^+$ toward Star $\iota$ (2MASS J17470898-2829561) and in spectra of H$_2$O$^+$, OH$^+$, and $^{13}$CH$^+$ toward Sgr B2(M) and Sgr B2(N) to measure the depths of these sources in the CMZ relative to Sgr~A$^\ast$. We find that each them lies $\sim$90 pc behind Sgr~A$^\ast$, near the rear of the CMZ. For the Sgr B2 sources, this conclusion contradicts many previous analyses \citep{rei09, kru15, rid17} in which Sgr B2 had been placed shallower than Sgr~A$^\ast$ in the CMZ.  Our conclusion agrees qualitatively with Figure 5 of \citet{mol11}. Note, however, that our face-on view of the CMZ is circular rather than an ellipse, with  Sgr~A$^\ast$ at the center of the circle. Our conclusion would be in good agreement with that of \citet[][Figure10]{sof95} if Arm I is the far spiral arm and Arm II is the near spiral arm, as proposed by \citet{rid17} and contrary to Sofue's assignment. 

\acknowledgments

We are grateful to the referee, Tetsuya Nagata, for a number of insightful comments  and for a suggestion regarding the $R$(2,2)$^l$ line toward Star $\iota$, which helped clarify our understanding of the chemistry in dense clouds.  We are indebted to Y. Sofue for useful discussions and for pointing out additional publications that consider the location of Sgr B2. We also thank P. Schilke for providing the far-infrared spectra in Figure 5, L. Armillotta for remarks on an earlier version of the manuscript, and M. J. Reid for a helpful conversation regarding trigonometric parallax. T.O. acknowledges a conversation with Harvey Liszt on the uncertainty in the location of the Sgr B2 complex. This paper is based in part on data obtained at the Gemini Observatory for programs GS-2009A-C-6 and GS-2010A-C-3. The Gemini Observatory is a program of NOIRLab,  which is managed by the Association of Universities for Research in Astronomy (AURA) under a cooperative agreement with the National Science Foundation, on behalf of the Gemini Observatory partnership: the National Science Foundation (United States), National Research Council (Canada), Agencia Nacional de Investigaci\'{o}n y Desarrollo (Chile), Ministerio de Ciencia, Tecnolog\'{i}a e Innovaci\'{o}n (Argentina), Minist\'{e}rio da Ci\^{e}ncia, Tecnologia, Inova\c{c}\~{o}es e Comunica\c{c}\~{o}es (Brazil), and Korea Astronomy and Space Science Institute (Republic of Korea). We thank the staffs of the Gemini Observatory for its support. The Herschel Telescope is a space observatory with scientific instruments provided by European-led Principal Investigator consortia and with important participation from NASA. This paper also is based in small part on data collected at the Subaru Telescope, which is operated by the National Astronomical Observatory of Japan. We are honored and grateful for the opportunity of observing the Universe from Maunakea, which has cultural, historical and natural significance in Hawaii. 
\\
\facilities{ Gemini:South, Subaru, Herschel Space Observatory}


\end{document}